# Mott transition in VO$_2$ revealed by infrared spectroscopy and nano-imaging

M. M. Qazilbash[1,*], M. Brehm[2], Byung-Gyu Chae[3], P.-C. Ho[1], G. O. Andreev[1], Bong-Jun Kim[3], Sun Jin Yun[3], A. V. Balatsky[4], M. B. Maple[1], F. Keilmann[2], Hyun-Tak Kim[3], D. N. Basov[1]

[1]*Physics Department, University of California-San Diego, La Jolla, California 92093, USA.*

[2]*Abt. Molekulare Strukturbiologie, Max-Planck-Institut für Biochemie & Center for NanoScience, 82152 Martinsried, München, Germany.*

[3]*IT Convergence & Components Lab, Electronics and Telecommunications Research Institute , Daejeon 305-350, Korea.*

[4]*Theoretical Division and Center for Integrated Nanotechnologies, MS B262, Los Alamos National Laboratory, Los Alamos, New Mexico 87545, USA.*

**To whom correspondence should be addressed. Email: mumtaz@physics.ucsd.edu*

## Abstract

Electrons in correlated insulators are prevented from conducting by Coulomb repulsion between them. When an insulator-to-metal transition is induced in a correlated insulator by doping or heating, the resulting conducting state can be radically different from that characterized by free electrons in conventional metals. We report on the electronic properties of a prototypical correlated insulator vanadium dioxide (VO$_2$) in which the metallic state can be induced by increasing temperature. Scanning near-field infrared microscopy allows us to directly image nano-scale metallic puddles that appear at the onset of the insulator-to-metal transition. In combination with far-field infrared spectroscopy, the data reveal the Mott transition with divergent quasiparticle mass in the metallic puddles. The experimental approach employed here sets the stage for investigations of charge dynamics on the nanoscale in other inhomogeneous correlated electron systems.

One challenge of contemporary condensed matter physics is the understanding of the emergence of metallic transport in correlated insulators or Mott insulators in which, for example, a temperature change or chemical doping induces anomalous conducting phases (*1*). In such a correlated metal the mobile charges experience strong competing interactions leading to exotic phases including the pseudogap state in cuprates and manganites, high-temperature superconductivity, charge stripes in cuprates, even phase separation in some manganites and cuprates (*1,2,3,4,5,6,7,8*). In systems where multiple phases coexist on the nanometer scale, the dynamical properties of these individual electronic phases remain unexplored because methods appropriate to study charge dynamics (transport, infrared/optical and many other spectroscopies) lack the required spatial resolution. Scanning near-field infrared microscopy can circumvent this limitation (*9,10,11*). Specifically, we probed coexisting phases in the vicinity of the insulator-to-metal transition in vanadium dioxide ($VO_2$) at length scales down to 20 nm. This enables us to identify an electronic characteristic of the Mott transition, namely divergent quasiparticle mass in the metallic puddles, which would otherwise have remained obscured in macroscopic studies that average over the coexisting phases in the insulator-to-metal transition regime.

One particular advantage of $VO_2$ for the study of electronic correlations is that the transition to the conducting state is initiated by increasing the temperature without the need to modify the stoichiometry. The salient features of the first-order phase transition that occurs at $T_c \approx 340$ K are the orders-of-magnitude increase in conductivity accompanied by a change in the lattice structure (*1*). Compared to the high temperature rutile metallic (R) phase, the two main features that distinguish the lattice in the low-temperature monoclinic (M1) insulating phase are dimerization (charge-ordering) of the vanadium ions into pairs, and the tilting of these pairs with respect to the c-axis of the rutile metal. The experiments on $VO_2$ films (*12,13*) reported here reveal a strongly correlated conducting state that exists within the insulator-to-metal transition region in the form of nanoscale metallic puddles. Electromagnetic response of these puddles separated by the insulating host displays the signatures of collective effects in the electronic system including divergent optical effective mass and optical pseudogap. These findings, which were not anticipated by theoretical models, may also help to settle the decades long debate (*1,14,15,16,17,18,19,20*) on the respective roles played by the lattice and by the electron-electron correlations in the insulator-to-metal transition.

The gross features of the insulator-to-metal transition in $VO_2$ can be readily identified through the evolution of the far-field optical constants (*13*) obtained using spectroscopic ellipsometry and reflectance (Fig. 1). The insulating monoclinic phase ($T \leq 341$ K) displays a sizable energy gap of about 4000 cm$^{-1}$ ($\approx 0.5$ eV) in the dissipative part of the optical conductivity $\sigma_1(\omega)$. The $T \geq 360$ K rutile metallic phase is characterized by a broad Drude-like feature in the optical conductivity, linear temperature dependence of resistivity, and an extremely short electronic mean free path of the order of the lattice constant, reminiscent of "bad metal" behavior in other transition metal oxides including the cuprates (*21,22,23*). The insulator-to-metal transition is evident from the increase of the conductivity with spectral weight "filling up" the energy gap that has to be contrasted with a gradual decrease of the energy gap magnitude. This feature of the transition, along with an isosbestic point at a frequency of $11500 \pm 125$ cm$^{-1}$, is one of several spectroscopic fingerprints of doped Mott insulators (*1*) identified in this work. The isosbestic point is defined here as the location of equal conductivity for all spectra obtained at different temperatures. Finally, the divergence of the real part of the dielectric function $\varepsilon_1$ (inset of Fig. 1) signals the percolative nature of the insulator-to-metal transition. This divergence of $\varepsilon_1$ is similar to that observed near the percolative insulator-to-metal transition in ultrathin Au and Pb films (*24*).

Mid-infrared near-field images directly show that in fact the insulating and metallic phases coexist in $VO_2$ over a finite temperature range in the transition region (see Fig. 2). This determination has been made using a scattering scanning near-field infrared microscope (s-SNIM) operating at the infrared frequencies $\omega = 930$ cm$^{-1}$ and $\omega = 1725$ cm$^{-1}$. S-SNIM is capable of registering contrast between electronic phases according to their optical constants with spatial resolution $\approx 20$ nm. Specifically, the scattering amplitude signal demodulated at the second harmonic of the tapping frequency of the tip of our s-SNIM apparatus (maps in Fig. 2) is related to the local value of the complex dielectric function $\widetilde{\varepsilon} = \varepsilon_1 + i\varepsilon_2$ of the sample. The amplitude of the scattering signal is expected to increase in metallic regions compared to the insulating regions: a behavior well-grasped by the so-called dipole model of the near-field infrared contrast (*9,10,13*).

The amplitude-contrast near-field images in Fig. 2 show the electronic insulator-to-metal transition in progress. At temperatures 295-341 K, in the insulating phase, we observe uniform maps of low scattering (dark blue color in Fig. 2). A small increase of temperature radically changes the near field images. For example, in the $T = 342.4$ K image we now observe nano-scale clusters where the amplitude of the scattering signal is enhanced by a factor of 2-5 compared to that of the insulating host indicating a metallic phase. Representative scans show that the metallic regions nucleate, then grow with increasing temperature and eventually connect. We do not observe any obvious correlations between the size and/or shape of the metallic clusters and the features in simultaneously collected topographic images. While the percolative nature of the insulator-to-metal transition had been proposed previously (*25*), it is directly revealed by our scanning near-field infrared measurements reported herein. The insulator-to-metal transition is complete by $T = 360$ K where insulating islands are no longer seen.

With the observation of nano-structured phases in Fig. 2, the far-field infrared spectra in Fig. 1 should be analyzed using an effective medium theory (EMT) for such phase-separated systems (*13,26*). The effective optical constants of a two-phase heterogeneous system are an average of the optical constants of the insulating and metallic regions weighted by the respective volume fractions. Our near-field images enable us to determine these fractions. However, a simple weighting of optical constants of the insulating phase and of the rutile metallic phase at $T = 360$ K within the EMT model does not produce a satisfactory description of the far-field infrared data near the onset of the insulator-to-metal transition in $VO_2$. This discrepancy indicates that the infrared properties of the metallic puddles, once they first appear at $T \approx 342$ K, may be different from that of the high temperature rutile metal. We confirmed this hypothesis by extracting the response of the metallic puddles from a combination of near field results and far field spectra within an EMT analysis described in the supporting online material (*13*).

The real part of the conductivity spectrum $\sigma_{1a}(\omega)$ of the metallic puddles is plotted in Fig. 3B as it evolves with temperature. When these puddles appear at the onset of the electronic insulator-to-metal transition at $T \approx 342$ K (see Fig. 3A), their conductivity spectrum differs markedly from that of the rutile metallic phase at higher temperature, for example, $T = 360$ K. These metallic regions exhibit a narrow Drude-like peak at low frequencies, and then a dip followed by a prominent mid-infrared band peaked at $\approx 1800$ cm$^{-1}$. Uncertainties in the EMT analysis (detailed in the supporting online material) do not exclude the possibility of a non-monotonic form of $\sigma_{1a}(\omega)$ at the lowest frequencies, a behavior consistent with Drude response modified by localization. These features indicate that the metallic islands are not simply isolated regions of the higher-temperature $VO_2$ rutile metal.

In order to highlight distinctions between the electrodynamics of the metallic clusters and the rutile metallic phase, we perform the extended Drude analysis (*13,27*) on the optical constants of the metallic clusters to extract the scattering rate $1/\tau(\omega)$ and the mass enhancement factor $m^*(\omega)/m_b$ ($m_b$ is the electronic band mass) of the charge carriers (see Fig. 3C,D). In the limit of $\omega \to 0$ these quantities can be interpreted in terms of lifetime $\tau(\omega)$ and effective mass $m^*(\omega)$ of quasiparticles (*27*) in the metallic regions. One can recognize a prominent enhancement of $m^*(\omega \to 0)/m_b$ at $T = 342$ K that has to be contrasted with much lighter masses in the rutile phase ($T = 360$ K) spectrum in Fig. 3**D** (*22*). More importantly, the temperature dependence of $m^*(\omega \to 0)/m_b$ plotted in the inset of Fig. 3D shows divergent behavior in the vicinity of the insulator-to-metal transition: an unambiguous attribute of the Mott transition (*28*). The spectra of $1/\tau(\omega)$ reveal a threshold structure followed by an overshoot at higher energies up to $\approx 1000$ cm$^{-1}$. This is characteristic of systems with a (pseudo)gap in the electronic density of states (*29*) that is to be contrasted with the relatively smooth variation of $1/\tau(\omega)$ in the rutile phase. We also note that the new electronic state exhibiting an enhanced mass and a gap-like form of the relaxation rate exists only in a narrow temperature range as shown by the shaded region in Fig. 3A. By $T = 343.6$ K, the optical constants of the metallic regions already resemble those of the rutile metallic phase.

The analysis and discussion above suggest that the classic temperature induced insulator-to-metal transition in $VO_2$ occurs from the monoclinic insulator to an incipient strongly correlated metal (SCM) in the form of nano-scale puddles. These metallic puddles exhibit mass divergence which is a clear signature of electronic correlations due to many-body Coulomb interactions (*28*). The pseudogap and mid-infrared band are the consequence of optically-induced electronic excitations across a gap on some parts of the Fermi surface. The energy scale of the pseudogap in the SCM state in $VO_2$ can be determined by the overshoot in $1/\tau(\omega)$ spectra that occurs at $\approx 1000$ cm$^{-1}$ (or $\approx 4k_B T_c$). We note that the pseudogap is a common property of doped Mott insulators (*1,27*). The pseudogap features in the optical conductivity and $1/\tau(\omega)$ spectra also bear resemblance to those found in metallic systems with partial charge density wave (CDW) gap (*30*). The pseudogap in $VO_2$ may result from a complex interplay between electronic correlations and charge ordering.

The Mott transition commonly leads to anti-ferromagnetically ordered insulator as in closely related $V_2O_3$ (*1*). Vanadium dioxide avoids this magnetic ordering via dimerization of vanadium ions in the monoclinic insulating phase (*14*) due to competing effects of charge ordering (Peierls instability) that is likely caused by electron-phonon interactions. Thus, the insulating monoclinic (M1) phase of $VO_2$ should be classified as a Mott insulator with charge-ordering. It remains an open question whether or not the insulator-to-metal transition occurs at a slightly different temperature from the structural transformation associated with charge ordering (*18,19,31*), and this raises the issue about the precise lattice structure of the metallic nano-puddles we have observed. This issue does not impact our observation of divergent optical mass, and can only be resolved by x-ray diffraction measurements on the nano-scale. We also note that the images of phase coexistence and percolation reported here (Fig. 2) are consistent with the thermodynamic evidence of the first order nature of the phase transition in $VO_2$ (*21*). Moreover, our experiments show that the collapse of a large $\approx 0.5$ eV energy gap and the formation of heavy quasiparticles in the emergent metallic nano-puddles at the onset of the insulator-to-metal transition are due to Mott physics (*1,28*), and percolation occurs at a later stage when these metallic puddles grow and connect (see Fig. 2).

A transformation from an insulator to a metal in many correlated electron systems including high-$T_c$ cuprates, colossal magnetoresistive manganites and others occurs through an intermediate pseudogap regime (*1,27,29*). At least in the case of cuprates, optical signatures of the pseudogap state are similar to the results in the SCM state of $VO_2$ (*27*). Furthermore, in many correlated systems, the pseudogap state is in the vicinity of the regime of the "bad metal" where resistivity shows a peculiar linear dependence with temperature whereas the absolute values of the resistivity are so large that the notion of quasiparticles becomes inapplicable (*21,22,23*). Often a crossover from pseudogapped metal to "bad metal" occupies an extended region of the phase diagram. In $VO_2$, the boundary between the two electronic regimes is relatively abrupt and the emergence of "bad metal" transport in the rutile phase may be linked to the loss of long-range charge order that does not extend into the rutile metal. Then the poor conductivity of rutile $VO_2$ and other "bad metals" appears to arise from the collapse of electronically and/or magnetically ordered states in the vicinity of a Mott insulator thereby causing the resistivity to exceed the Ioffe-Regel-Mott limit of metallic transport (*21,22,23*). Finally, we note here that in the cuprates, in contrast to $VO_2$, the effective mass of doped carriers inferred from infrared spectroscopy data (*32*) shows no divergence. However, if electronic phase separation exists in doped cuprates as suggested by recent scanning probe studies (*6,8*), then infrared analysis of the effective quasiparticle mass needs to be revisited with the help of nano-imaging tools employed in this work.

# References

1. M. Imada, A. Fujimori, Y. Tokura, *Rev. Mod. Phys.* **70**, 1039 (1998).

2. V. J. Emery, S. A. Kivelson, J. M. Tranquada, *Proc. Nat. Acad. Sci USA* **96**, 8814 (1999).

3. M. Uehara, S. Mori, C. H. Chen, S.-W. Cheong, *Nature* **399**, 560 (1999).

4. S. Liuwan, C. Israel, A. Biswas, R. L. Greene, A. de Lozanne, *Science* **298**, 805 (2002).

5. E. Dagotto, *Science* **309**, 257 (2005).

6. J. Lee *et al.*, *Nature* **442**, 546 (2006).

7. Z. Sun *et al.*, *Nat. Phys.* **3**, 248 (2007).

8. K. K. Gomes *et al.*, *Nature* **447**, 569 (2007).

9. B. Knoll, F. Keilmann, *Nature* **399**, 134 (1999).

10. F. Keilmann, R. Hillenbrand, *Phil. Trans. R. Soc. Lond. A* **362**, 787 (2004).

11. N. Ocelic, A. Huber, R. Hillenbrand, *App. Phys. Lett.* **89**, 101124 (2006).

12. B. G. Chae *et al.*, *Electrochem. Solid-State Lett.* **9**, C12 (2006).

13. See Supporting Online Material.

14. R. M. Wentzcovitch, W. W. Schulz, P. B. Allen, *Phys. Rev. Lett.* **72**, 3389 (1994).

15. T. M. Rice, H. Launois, J. P. Pouget, *Phys. Rev. Lett.* **73**, 3042 (1994).

16. S. Biermann, A. Poteryaev, A. I. Lichtenstein, A. Georges, *Phys. Rev. Lett.* **94**, 026404 (2005).

17. T. C. Koethe *et al.*, *Phys. Rev. Lett.* **97**, 116402 (2006).

18. H. T. Kim *et al.*, *Phys. Rev. Lett.* **97**, 266401 (2006).

19. H.-T Kim *et al.*, *New Journal of Physics* **6**, 52 (2004).

20. E. Arcangeletti *et al.*, *Phys. Rev. Lett.* **98**, 196406 (2007).

21. P. B. Allen, R. M. Wentzcovitch, W. W. Schulz, P. C. Canfield, *Phys. Rev. B* **48**, 4359 (1993).

22. M. M. Qazilbash *et al.*, *Phys. Rev. B* **74**, 205118 (2006).

23. V. J. Emery, S. A. Kivelson, *Phys. Rev. Lett.* **74**, 3253 (1995).

24. J. J. Tu, C. C. Homes, M. Strongin, *Phys. Rev. Lett.* **90**, 017402 (2003).

25. H. S. Choi, J. S. Ahn, J. H. Jung, T. W. Noh, D. H. Kim, *Phys. Rev. B* **54**, 4621 (1996).

26. G. L. Carr, S. Perkowitz, D. B. Tanner, Far infrared properties of inhomogeneous materials, *Infrared and millimeter waves, vol. 13*, edited by Kenneth J. Button (Academic Press, Orlando, 1985).

27. D. N. Basov, T. Timusk, *Rev. Mod. Phys.* **77**, 721 (2005).

28. W. F. Brinkman, T. M. Rice, *Phys. Rev. B* **2**, 4302 (1970).

29. D. N. Basov, E. J. Singley, S. V. Dordevic, *Phys. Rev. B* **65**, 054516 (2002).

30. G. Gruner, *Density waves in solids* (Perseus Publishing, Cambridge, Massachusetts, 2000).

31. Y. J. Chang *et. al.*, *Phys. Rev. B* **76**, 075118 (2007).

32. W. J. Padilla *et. al.*, *Phys. Rev. B* **72**, 060511(R) (2005).

33. This work was supported by the US Department of Energy, the DFG Cluster of Excellence Munich-Centre for Advanced Photonics, and the High Risk High Return project in the Electronics and Telecommunications Research Institute (ETRI), Korea.

**Figure Legends**

**Fig. 1**: The real part of the optical conductivity $\sigma_1(\omega) = \frac{\omega\varepsilon_2(\omega)}{4\pi}$ of VO$_2$ is plotted as a function of frequency for various representative temperatures. The open circle denotes the isosbestic (equal conductivity) point for all spectra. The inset shows the temperature dependence of the real part of the dielectric function $\varepsilon_i$ in the low frequency limit ($\omega = 50$ cm$^{-1}$).

**Fig. 2**: Images of the near-field scattering amplitude over the same 4 µm x 4 µm area obtained by scattering scanning near-field infrared microscope (s-SNIM) operating at the infrared frequency $\omega = 930$ cm$^{-1}$. These images are displayed for representative temperatures in the insulator-to-metal transition regime of VO$_2$ to show percolation in progress. The metallic regions (light blue, green and red colors) give higher scattering near-field amplitude compared to the insulating phase (dark blue color). See supporting online material for details (*13*).

**Fig. 3**: (**A**) The phase diagram of VO$_2$ and the resistance-temperature curve showing the insulator-to-metal transition. The shaded area highlights the region of the phase diagram in which the strongly correlated metal (SCM) with divergent quasiparticle mass and an optical pseudogap exists. Panels (**B**), (**C**), and (**D**) show the evolution of the optical conductivity $\sigma_{1a}(\omega)$, the scattering rate $1/\tau(\omega)$, and the optical effective mass normalized by the band value $m^*(\omega)/m_b$ of the metallic regions of VO$_2$ with increasing temperature. The inset in panel (**D**) shows the $\omega \rightarrow 0$ limit of the mass enhancement factor as a function of temperature. The data points between $T = 400$ K and 550 K are taken from Ref.(*22*).

**Fig.1**

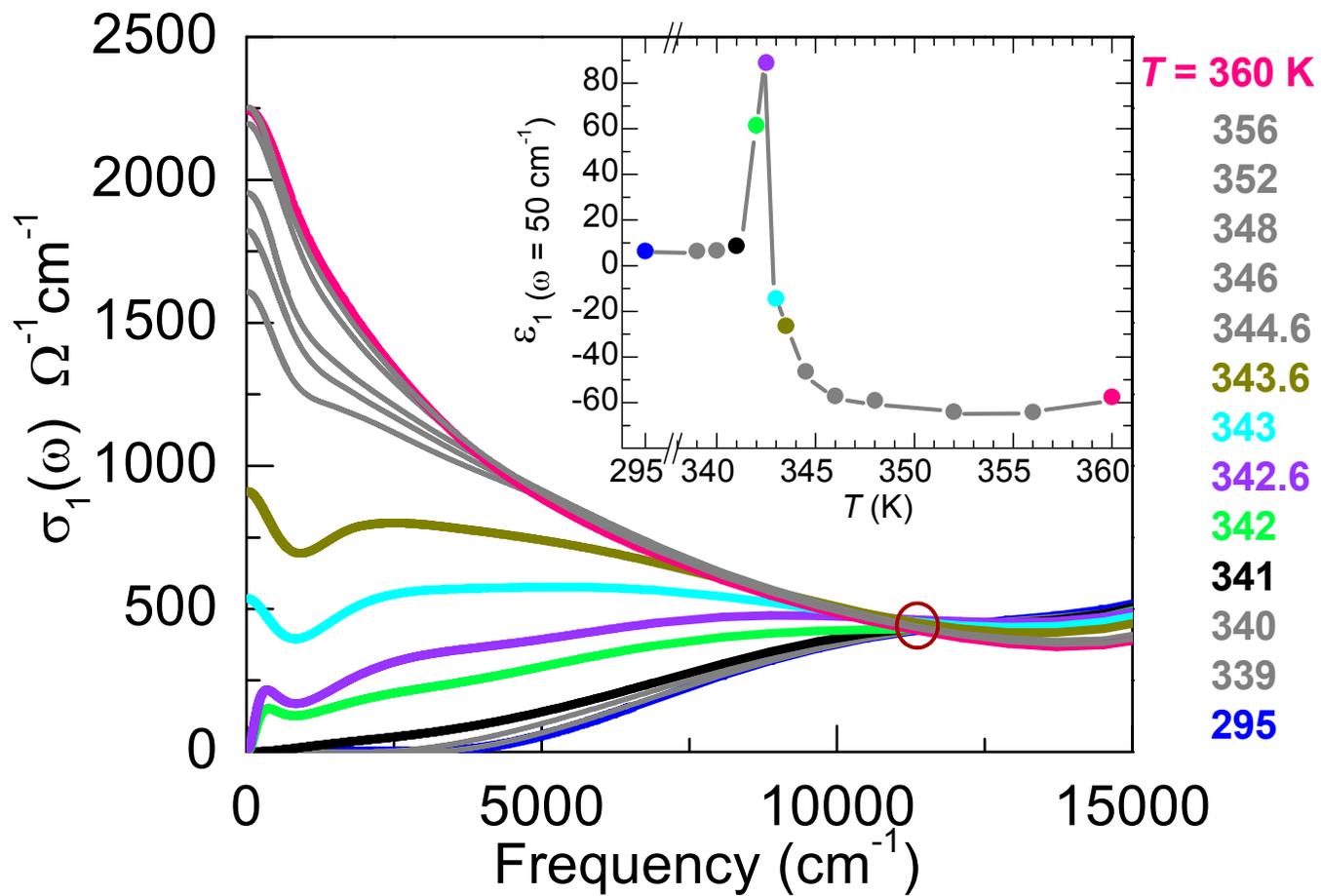

$T$ = 360 K
356
352
348
346
344.6
343.6
343
342.6
342
341
340
339
295

**Fig.2**

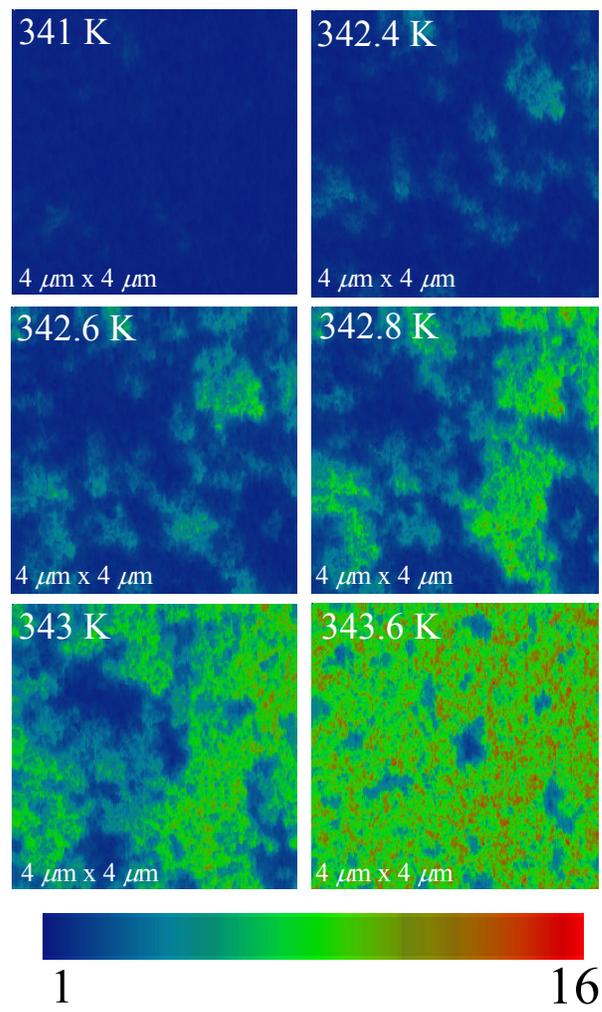

**Fig.3**

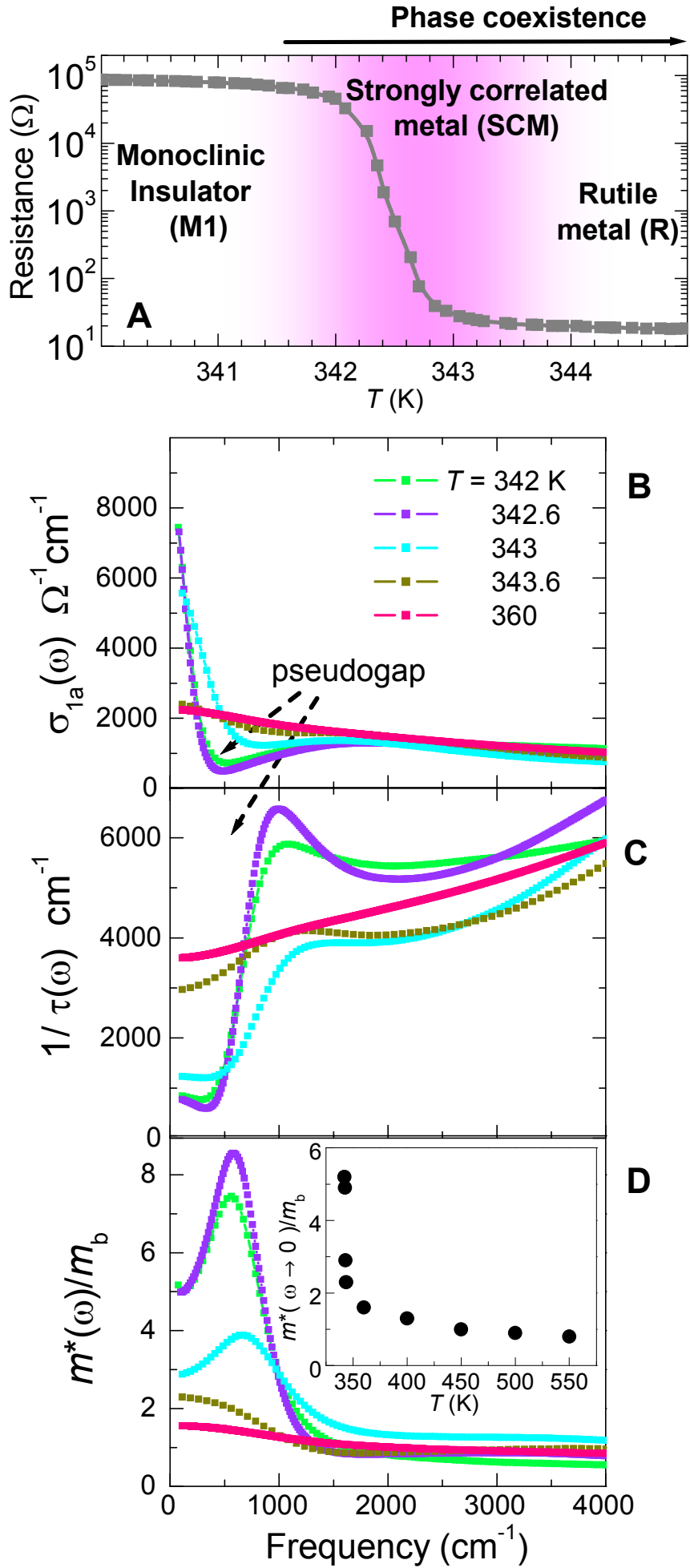

## Supporting Online Material

### Samples and measurements

The VO$_2$ films used in this study are $\approx$ 100 nm thick and were grown on $(\overline{1}012)$ oriented Al$_2$O$_3$ (sapphire) substrates. Details of film growth and characterization are given in Ref.(*S1*). We chose thin films of VO$_2$ for infrared measurements in the phase transition regime instead of single crystals because the crystals fracture when taken across the phase transition and thereby create problems of reproducibility and significant uncertainties in the quantitative results in the transition regime. Data obtained for high quality films grown on lattice matched substrates are free of the above complications and thus enable studies of the intrinsic properties of VO$_2$.

The data and results presented in this report were obtained while heating up the sample. The data and results for the cooling runs show the same behavior as obtained for the heating runs for all the measurements but in reverse order. There is also a downward shift in $T_c$ by $\approx$ 8 K on cooling compared to the $T_c$ on heating due to hysteresis resulting from the first order nature of the phase transition (see Ref. *S1*).

The resistance of the film plotted in Fig. 3a was measured by a standard four-probe method. The technique of scanning near-field infrared microscopy (*S2,S3,S4*) is described in the following section. The far-field complex optical constants (or dielectric function) $\widetilde{\varepsilon}(\omega) = \varepsilon_1(\omega) + i\varepsilon_2(\omega)$ were obtained via a combination of ellipsometry in the spectral range 400 - 20000 cm$^{-1}$ and near-normal incidence reflectance in the spectral range 40 - 680 cm$^{-1}$ (see below for details). We note here that in general the complex conductivity $\widetilde{\sigma}(\omega) = \sigma_1(\omega) + i\sigma_2(\omega)$ of a material is related to its complex optical (dielectric) constant $\widetilde{\varepsilon}(\omega)$ by $\widetilde{\varepsilon}(\omega) = 1 + \dfrac{4\pi i}{\omega}\widetilde{\sigma}(\omega)$. The effective medium theory (EMT) was employed to obtain the optical constants of the metallic puddles in the insulator-to-metal transition regime (*S5,S6*). The details of the EMT and the extended Drude analysis (*S7*) are given in subsequent sections.

**Scanning near-field infrared microscopy.**  Scattering scanning near-field infrared microscopy was performed with a custom Atomic Force Microscope (AFM) using Pt-coated silicon tips in the tapping mode (*S2,S3*). The $\omega = 930$ cm$^{-1}$ and $\omega = 1725$ cm$^{-1}$ infrared frequencies were provided by a $CO_2$ laser and a CO laser respectively. The pseudo-heterodyne scheme with second harmonic demodulation was employed to isolate and detect the near-field signal in amplitude and phase (*S4*). The spatial resolution of the near-field infrared probe is about 10-20 nm and is set by the radius of curvature of the tip. The near-field penetrates about 20 nm into the $VO_2$ film and is a bulk probe of the insulator-to-metal transition compared to the scanning tunneling microscope, for example, which is a surface sensitive probe.

The near-field interaction between the tip and the sample is described by the dipole model. Within this model, the complex scattering signal at the second harmonic of the tapping frequency is a function of the optical constants of the tip and the sample (*S2,S3*). The scattering amplitude is higher for metallic regions which have large negative real part and large positive imaginary part of the optical constants at $\omega = 930$ cm$^{-1}$ and $\omega = 1725$ cm$^{-1}$. Insulators like monoclinic $VO_2$ yield lower scattering amplitudes because of the negligibly small imaginary part of the optical constants and a positive real part somewhat higher than unity ($\varepsilon_1$ [$\omega = 930$ cm$^{-1}$]$\approx 5$ and $\varepsilon_1$ [$\omega = 1725$ cm$^{-1}$] $\approx 6$ in insulating $VO_2$). This difference in scattering amplitudes allows us to image the insulator-to-metal transition in $VO_2$. The near-field scattering amplitude images obtained with $\omega = 1725$ cm$^{-1}$ are similar to those shown in Fig. 2 which were obtained with $\omega = 930$ cm$^{-1}$.

The dipole model provides a good quantitative description of the near-field infrared scattering amplitudes near the onset of the insulator-to-metal transition where the metallic puddles are separated by the insulating host. The model also captures the observed near-field contrast at higher fractions of the metallic regions but will need to be refined to quantify the continuously increasing scattering amplitude from the expanding metallic regions. However, this does not impact the determination of the insulating and metallic fractions from near-field images.

We note that repeated near-field scans in the insulator-to-metal transition regime over the same sample area and at a fixed temperature show nearly identical patterns of metallic puddles in the insulating host. This indicates that the effect of possible dynamic fluctuations on the static patterns is small. The metallic islands are most likely seeded and controlled by stress at the interface, defects, grain boundaries etc. and this is a reasonable scenario for realistic samples. These nucleation centers would lead to static and reproducible patterns of inhomogeneity at a given temperature. Therefore, the fraction of the static metallic (and insulating) regions obtained from near-field images can be used to interpret far-field data within the effective medium theory described in a subsequent section.

**Far-field infrared spectroscopy.** Spectroscopic ellipsometry provides the real and imaginary parts of the ellipsometric coefficients at each measured frequency (*S8*). This enables us to obtain the optical constants of the $VO_2$ film without recourse to Kramers-Kronig analysis which breaks down in the insulator-to-metal transition region of $VO_2$ because of inhomogeneity and phase-coexistence. Provided the incident wavelength is large compared to the size of the nano-scale metallic puddles, the inhomogeneous system can be described by an effective dielectric function (*S5,S6*). This is certainly expected to be the case for long wavelengths in the far- and mid-infrared regime (between 40 cm$^{-1}$ and 5000 cm$^{-1}$). In accord with the Bruggeman effective medium theory (EMT) (*S5,S6*), simple mixing of the dielectric functions of the insulating phase and rutile metallic phase ($T$ = 360 K) in the mid-infrared to visible regime (3000-20000 cm$^{-1}$) provides a good description of the data. Therefore, it is appropriate to assign an effective dielectric function to the inhomogeneous, phase-separated regime of VO$_2$ even when the wavelength becomes comparable to the size of the metallic puddles.

The effective dielectric function $\widetilde{\varepsilon}_E(\omega)$ of the inhomogeneous, phase-separated regime of VO$_2$ is thus obtained via the standard analysis of the combined ellipsometric and reflectance data based on a two-layer model of a VO$_2$ film with an effective dielectric function on a sapphire substrate (*S8*). The contributions of VO$_2$ phonons to $\widetilde{\varepsilon}_E(\omega)$ were modeled by Lorentzian oscillators and subsequently the phonon contributions were subtracted so that the electronic contribution to the optical constants could be unambiguously presented and analyzed.

**Effective medium theory.** A simple weighting of optical constants of the insulating phase and of the metallic phase ($T = 360$ K spectrum) within the effective medium theory (EMT) produces a good description of the far-field infrared data near the onset of the insulator-to-metal transition at frequencies between 3000 cm$^{-1}$ and 20000 cm$^{-1}$. Deviations from this description become progressively stronger at longer wavelengths (40 to 3000 cm$^{-1}$), a regime where the EMT formalism is expected to be most directly applicable (*S5,S6*). This implies that the optical constants of the metallic puddles when these first appear at the onset of the insulator-to-metal transition are different from the optical constants of the rutile metal ($T = 360$ K). Therefore, we use the EMT to extract the optical constants of the metallic puddles.

Within the Bruggeman effective medium theory (*S5,S6*), the effective (complex) optical constants $\widetilde{\varepsilon}_E(\omega)$ of a two-component inhomogeneous system are given by:

$$f \frac{\widetilde{\varepsilon}_a(\omega) - \widetilde{\varepsilon}_E(\omega)}{\widetilde{\varepsilon}_a(\omega) + \frac{(1-q)}{q}\widetilde{\varepsilon}_E(\omega)} + (1-f)\frac{\widetilde{\varepsilon}_b(\omega) - \widetilde{\varepsilon}_E(\omega)}{\widetilde{\varepsilon}_b(\omega) + \frac{(1-q)}{q}\widetilde{\varepsilon}_E(\omega)} = 0 \qquad (1)$$

In this equation, $\widetilde{\varepsilon}_a(\omega)$ and $\widetilde{\varepsilon}_b(\omega)$ are the complex optical constants of the metallic and insulating phases respectively, $f$ and $(1-f)$ are the volume fractions of the metallic and insulating phases respectively, and $q$ is the depolarization factor that depends on the shape of the components which is inferred from near-field images. This factor is taken to be 0.2-0.4 assuming nearly spherical metallic regions at low concentrations of the newborn metallic state because their lateral dimensions are less than the film thickness and the out-of-plane dimension is assumed to be similar to the lateral dimensions. The depolarization factor continuously increases to 0.5 which is appropriate for thin flat disks at higher concentrations of metallic clusters as their lateral dimensions exceed the out-of-plane dimension which is limited by the thickness of the film (*S6*). The volume fractions of the insulating and metallic phases are obtained directly from the near-field images. The far-field optical constants of the insulating phase $\widetilde{\varepsilon}_b(\omega)$ and the inhomogeneous transition regime $\widetilde{\varepsilon}_E(\omega)$ have also been measured (see Fig. 1). The optical constants of the insulating phase are assumed to be independent of temperature in this analysis. This is supported by the near-field images which show little variation of the scattering intensity from the insulating regions with increasing temperature. Since $\widetilde{\varepsilon}_E(\omega)$, $\widetilde{\varepsilon}_b(\omega)$, and $f$ are known whereas the range of $q$ is constrained based on the knowledge of the metallic cluster size and film thickness, the equation can be solved to obtain the optical constants of the metallic regions $\widetilde{\varepsilon}_a(\omega)$, and hence the complex conductivity $\widetilde{\sigma}_a(\omega)$ whose real part $\sigma_{1a}(\omega)$ is plotted in Fig.3B. Values of $f$ and $q$ at relevant temperatures are listed in Table S1. Uncertainties in the magnitude of $q$ do affect the behavior of $\widetilde{\sigma}_a(\omega)$ spectra at the lowest frequencies close to the experimental cut-off of the data. Specifically, these uncertainties do not rule out a non-monotonic form of the conductivity $\sigma_{1a}(\omega)$: a response consistent with Drude dynamics modified by localization. Irrespective of some ambiguity in the form of the conductivity in the limit of $\omega \to 0$, the oscillator strength of this low-energy mode is continuously decreasing as the transition to the insulator is produced, signaling an increase of the optical effective mass.

Interfacial scattering and localization will become important when the size of the metallic puddles is comparable to or less than the intrinsic mean free path of the charge carriers within these puddles. Our estimate of the mean free path based on the scattering rate data in Fig.3C is of the order of few nanometers. Thus the smallest puddles we can detect (limited by the 20 nm spatial resolution of our probe) are still in the regime where the mean free path is shorter than the spatial extent of metallic regions. Therefore, we believe interfacial scattering and localization are unlikely to affect charge dynamics in the metallic puddles. However, we cannot rule out signatures of localization below our experimental cut-off frequency.

**Extended Drude formalism.** The following equations of the extended Drude formalism (*S7*) completely describe the evaluation of the scattering rate $1/\tau(\omega)$ and the mass enhancement factor $m^*(\omega)/m_b$ from the real and imaginary parts of the optical conductivity of the metallic regions $\tilde{\sigma}_a(\omega) = \sigma_{1a}(\omega) + i\sigma_{2a}(\omega)$ :

$$\frac{1}{\tau(\omega)} = \left(\frac{\omega_p^2}{4\pi}\right) \frac{\sigma_{1a}(\omega)}{\sigma_{1a}^2(\omega) + \sigma_{2a}^2(\omega)} \qquad (2)$$

$$\frac{m^*(\omega)}{m_b} = \left(\frac{\omega_p^2}{4\pi\omega}\right)\frac{\sigma_{2a}(\omega)}{\sigma_{1a}^2(\omega) + \sigma_{2a}^2(\omega)} \qquad (3)$$

The plasma frequency $\omega_p = 22000$ cm$^{-1}$ is nearly constant (within ten percent) for the metallic states and is obtained via the partial sum rule:

$$\frac{\omega_p^2}{8} = \int_0^{\omega_c} \sigma_{1a}(\omega)d\omega \qquad (4)$$

Here, the frequency cut-off $\omega_c = 13700$ cm$^{-1}$ is chosen to exclude contributions from higher-lying optical transitions.

The plasma frequency is related to the carrier density ($n$) by the following expression:

$$\omega_p^2 = \frac{4\pi ne^2}{m_b} \qquad (5)$$

Since the experimentally determined plasma frequency is nearly constant for the metallic states, we infer that the carrier density is also nearly constant for the evolving metallic regions provided that the band mass ($m_b$) of the carriers does not change drastically.

| $T$ (K) | $f$ | $q$ |
|---|---|---|
| 342 | 0.18 | 0.2 |
| 342.6 | 0.31 | 0.33 |
| 343 | 0.48 | 0.45 |
| 343.6 | 0.7 | 0.5 |

Table S1. The values of the filling fraction $f$ and depolarization factor $q$ used to obtain the optical constants of the metallic regions at selected temperatures within the framework of the EMT formalism (Eq.1).

*S1*. B. G. Chae *et al.*, *Electrochem. Solid-State Lett.* **9**, C12 (2006).

*S2*. B. Knoll, F. Keilmann, *Nature* **399**, 134 (1999).

*S3*. F. Keilmann, R. Hillenbrand, *Phil. Trans. R. Soc. Lond. A* **362**, 787 (2004).

*S4*. N. Ocelic, A. Huber, R. Hillenbrand, *App. Phys. Lett.* **89**, 101124 (2006).

*S5*. K. D. Cummings, J. C. Garland, D. B. Tanner, *Phys. Rev. B* **30**, 4170 (1984).

*S6*. G. L. Carr, S. Perkowitz, D. B. Tanner, Far infrared properties of inhomogeneous materials, *Infrared and millimeter waves, vol. 13*, edited by Kenneth J. Button (Academic Press, Orlando, 1985).

*S7*. D. N. Basov, T. Timusk, *Rev. Mod. Phys.* **77**, 721 (2005).

*S8*. H. G. Tompkins, E. A. Irene, editors, *Handbook of Ellipsometry*, William Andrew Publishing/Springer Verlag (2005).